\documentclass[prb,twocolumn,superscriptaddress,showpacs,amsmath,amssymb,longbibliography]{revtex4-2}
\usepackage{graphicx, color}
\usepackage{dcolumn}
\usepackage{bm}
\usepackage{epstopdf}
\usepackage{physics}
\usepackage[usenames,dvipsnames]{xcolor}
\usepackage{subfigure}
\usepackage{pifont}

\definecolor{orange}{rgb}{1,0.5,0}
\definecolor{goodgreen}{rgb}{0.1,0.5,0}
\definecolor{goodred}{rgb}{0.7,0,0}
\usepackage[colorlinks,urlcolor=goodgreen,citecolor=blue,linkcolor=goodred]{hyperref}
\usepackage[normalem]{ulem}

\usepackage{verbatim}
\graphicspath{ {./images/} }

\let\oldepsilon\epsilon \let\epsilon\varepsilon \let\varepsilon\oldepsilon

\let\Re\myRe

\let\Im\myIm

\begin{document}
\title{Dynamical Hall responses of disordered superconductors}

\begin{abstract}
We extend the Mattis-Bardeen theory for the dynamical response of superconductors to include different types of Hall responses. This is possible thanks to a recent  modification of the quasiclassical Usadel equation, which allows for analyzing Hall effects in disordered superconductors and including the precise frequency dependence of such effects. Our results form a basis for analyzing dynamical experiments especially on novel thin-film superconductors, where ordinary Hall and spin Hall effects can both show up.
\end{abstract}

\author{Alberto Hijano}
\email{alberto.hijano@ehu.eus}
\affiliation{Centro de F\'isica de Materiales (CFM-MPC) Centro Mixto CSIC-UPV/EHU, E-20018 Donostia-San Sebasti\'an, Spain}
\affiliation{Department of Condensed Matter Physics, University of the Basque Country UPV/EHU, 48080 Bilbao, Spain}

\author{Sakineh Vosoughi-nia}
\email{sakineh.vosooghi@gmail.com}
\affiliation{Department of Physics and Nanoscience Center, University of Jyväskylä, P.O. Box 35 (YFL), FI-40014 University of Jyvaskyla, Finland}
\affiliation{AGH University of Krakow, Academic Centre for Materials and Nanotechnology, al. A. Mickiewicza 30, 30-059 Krakow, Poland}
\affiliation{Institute of Physics, M. Curie-Skłodowska University, 20-031 Lublin, Poland}

\author{F. Sebasti\'an Bergeret}
\affiliation{Centro de F\'isica de Materiales (CFM-MPC) Centro Mixto CSIC-UPV/EHU, E-20018 Donostia-San Sebasti\'an,  Spain}
\affiliation{Donostia International Physics Center (DIPC), 20018 Donostia--San Sebasti\'an, Spain}

 \author{Pauli Virtanen}
\affiliation{Department of Physics and Nanoscience Center, University of Jyväskylä, P.O. Box 35 (YFL), FI-40014 University of Jyvaskyla, Finland}

 \author{Tero T. Heikkil\"a}
 \email{tero.t.heikkila@jyu.fi}
\affiliation{Department of Physics and Nanoscience Center, University of Jyväskylä, P.O. Box 35 (YFL), FI-40014 University of Jyvaskyla, Finland}

\maketitle

\section{Introduction}
Simultaneous application of electric and magnetic fields  on a conductor leads to the presence of a charge current with a transverse component perpendicular to both fields, in addition to the ordinary longitudinal current in the direction of the electric field. This ordinary Hall effect has been known since the 19th century~\cite{hall1879new} and it can be directly incorporated into the Drude model \cite{drude1900elektronentheorie,drude1900elektronentheorie2} of electronic conduction once the Lorenz force due to the magnetic field is included. Varying the electric field in time leads to similarly varying longitudinal and transverse charge currents~\cite{Ashcroft-Mermin}. This dynamical Hall effect can be observed for example in optical spectroscopy via the Faraday-Kerr rotation of the polarization state of light \cite{Magneto-Optics,Ebert:1996}. For frequencies low compared to the scattering rate and for materials in their normal state, both longitudinal and Hall currents are in phase with the electric field. This is in contrast with the superconducting state~\cite{mattis1958theory}, featuring both in-phase and out-of-phase contributions. For the longitudinal response, the former describes electronic transitions and features a superconducting gap at low temperatures, whereas the latter results from the supercurrent. Despite some attempts over the years \cite{miller1961frequency,Spielman:1994} based on phenomenological two-fluid models and Bardeen-Cooper-Schrieffer (BCS) theory,
to our knowledge the microscopic extension of the Drude model for the dynamical Hall response in superconductors in the dirty limit has not been presented before. We fill this gap by deriving the frequency dependent linear conductivity of dirty superconductors in the presence of the Hall effect and discuss how the in- and out-of-phase parts of the transverse response show up in the amplitude and phase of the frequency dependent Kerr response of such materials.

In type II superconductors, motion of vortices and the flux they carry gives additional contributions to the Hall effect~\cite{Bardeen:1965,Blatter:1994,Kopnin:2001,Kopnin:2002}. By now, especially in the steady state, these effects are well studied. Here we consider situations below the critical field in which no vortices are present, concentrating on the time-dependent response in the uniform gapped state.

In the presence of spin-orbit interaction, another type of  Hall effect called the spin Hall effect occurs~\cite{sinova2015spin}. It involves the generation of a transverse spin current in response to a charge current. There are two major mechanisms for this spin Hall effect: in the {\it intrinsic} mechanism, it is produced by the inversion symmetry breaking either due to the lattice (Dresselhaus spin-orbit coupling (SOC)~\cite{Dresselhaus}) or the sample structure (Rashba spin-orbit coupling~\cite{Rashba}) and in the {\it extrinsic} mechanism it results from the spin-dependence of the scattering.

In superconductors, the spin Hall effect couples (equilibrium) supercurrents and spin~\cite{edelstein1995magnetoelectric,yip2002two,dimitrova2007theory,agterberg2012magnetoelectric,konschelle2015theory}.
In addition, superconductors show also a quasiparticle spin Hall effect, which behaves otherwise similar to the normal-state version but depends strongly on temperature. Vortex motion can also generate it~\cite{Vargunin:2019}.
In this paper, we examine the dynamical spin Hall response in superconductors and show how it also contains in- and out-of-phase contributions similar to the longitudinal current response. In the normal state, our frequency dependent results are consistent with the literature predictions~\cite{Mishchenko:2004,Gorini:2012}, where the intrinsic spin Hall current is maximal at frequencies comparable with the spin-relaxation rate, whereas the extrinsic mechanism produces spin Hall response also at low frequencies. Superconductivity actually provides a tool for probing these different mechanisms as in the intrinsic case it leads to strongly temperature dependent spin Hall responses. Hence, whereas the frequency dependence may be difficult to probe on a wide scale of the order of the spin-relaxation rate, in the superconducting case one may fix the frequency and rather vary the temperature. Such studies may then provide information about the nature of the relevant mechanisms for the spin Hall effect.

Our work is based on the recent extensions of the quasiclassical Usadel equation to govern ordinary and spin Hall effects~\cite{Virtanen:2021,Virtanen:2022}, both in the extrinsic and intrinsic cases. We utilize these extensions here to study those dynamic responses. 
The dynamical Hall effect can be studied on conventional  spin singlet superconductors in the presence of the time-independent magnetic field. Because of Meissner screening, it shows up as a surface effect, but the same is true for the normal state because of the finite skin depth. On the other hand, the spin Hall effects require strong spin-orbit interaction and are especially interesting in thin-film systems involving either heavy-metal superconductors or the presence of a nearby heavy metal in which case the spin-orbit coupling would enter as a proximity effect. On the other hand, our work provides a baseline to compare the results of dynamical experiments on frequency dependent electromagnetic susceptibility of two-dimensional superconductors where possible spin ordering or orbital degrees of freedom may complicate the dynamic response. 

Our paper is organized as follows. In Sec.~\ref{sec:theory} we outline the theory for describing the various Hall effects in superconductors by introducing ordinary and SU(2) vector potentials and the accompanying field strength tensor terms into the Usadel equation. In Sec.~\ref{sec:onsager} we analyze the symmetry properties of the resulting dynamical response matrix. This section uses dynamical SU(2) fields as a formal tool for uncovering those symmetries. Section \ref{sec:Mattis-Bardeen} shows how the ordinary Mattis-Bardeen response \cite{mattis1958theory} naturally comes from our formalism. Then Sec.~\ref{sec:hall} discusses the dynamical Hall response, and Secs.~\ref{sec:spin Hall} and \ref{sec:Inverse spin Hall response} the dynamical spin Hall and inverse spin Hall responses in superconductors. Finally, Sec.~\ref{sec:conclusions} discusses the results and possible extensions of the theory to new materials. 

\section{Theory of  Hall effects in disordered superconductors}
\label{sec:theory}

In this section, we introduce the specific scenarios and main equations used  in this paper. Our study encompasses a broad scope, focusing on superconductors subjected to electromagnetic fields, alongside exchange fields and linear-in-momentum spin-orbit coupling, which can be treated as effective SU(2) potentials~\cite{Bergeret:2013-SU2}. These can be described by the following Hamiltonian:
\begin{equation}
    \mathcal{H}=\frac{(\boldsymbol{p}-\check{\boldsymbol{A}})^2}{2m}-\mu_\mathrm{ch}+V_{\mathrm{imp}}+\tau_3 \check{A}^0-i\check{\Delta}\; ,
\end{equation}
where $\boldsymbol{p}$ is the momentum, $m$ is the electron mass, $\mu_\mathrm{ch}$ is the chemical potential, $\check{\Delta}=\Delta\tau_1$ is the superconducting order parameter for s-wave superconductors, and $\sigma_j$ and $\tau_j$ are the Pauli matrices in spin and Nambu spaces, respectively. $V_{\mathrm{imp}}$ is a random impurity potential 
that consists of the usual elastic scattering  and the spin-orbit interaction~\cite{Bergeret:2016,Huang:2018,Virtanen:2021}. 
$\check{A}^\mu$ is the generalized  four-potential containing both U(1), and  SU(2) components~\cite{Bergeret:2014-SU2,Bergeret:2015-SU2,Tokatly:2017} given by 
\begin{subequations}\label{eq:fourpotential}
\begin{align}
    \check{A}^0 &= -e\phi\tau_3+\frac{\hbar}{2}A^{0j}\sigma_j\\
    \check{A}^i &= -eA^i\tau_3+\frac{\hbar}{2}A^{ij}\sigma_j\; .
\end{align}
\end{subequations}
$\phi$ and $\boldsymbol{A}$ are the usual U(1) scalar and vector electromagnetic potentials, while $A^{0j}$ and $A^{ij}$ are SU(2) potentials describing the Zeeman or exchange field and the linear-in-momentum SOC, respectively~\cite{Gorini:2012}. Here and below sum over repeated indices is assumed.

As in conventional electrodynamics we 
can define the field strength associated with $\check{A}$:  \begin{equation}\label{field strength}
    \check{F}^{\mu\nu}=\partial^\mu \check{A}^{\nu}-\partial^\nu \check{A}^{\mu}-\frac{i}{\hbar}[\check{A}^{\mu},\check{A}^{\nu}]\; .
\end{equation}
The last commutator appears because of  the fact that the SU(2) components are non-abelian. Here and below Greek indices range $\mu=0,1,2,3$.

In what follows, we are interested in the currents (charge and spin) generated by  the electric field, which is given by the $\check{E}^{j}\equiv\check{F}^{0j}=-eE^{j}\tau_3+(\hbar/2)E^{jl}\sigma_l$ components of the field strength tensor~\eqref{field strength}, where the Latin indices range $j=1,2,3$. 
In the linear response regime the  current and the field 
are related via the response tensor:
\begin{equation}\label{Ohms law}
    j^{i\mu}(\omega)=\sigma^{i\mu,j\nu}(\omega)E^{j\nu}(\omega).
\end{equation}
Here $j^{i0}$ are the components of the charge current whereas $j^{ij}$ is the spin-current tensor. The usual U(1) electric  field is given by $E^j \equiv E^{j0}$, and $E^{jl}$ denote the components of the SU(2) electric field.    The real part of $\sigma^{i\mu,j\nu}$ describes the in-phase response, and the imaginary part is the out-of-phase response of the current to the field. 

Our goal  is to find the conductivity tensor $\sigma^{i\mu,j\nu}(\omega)$ in  diffusive superconducting systems showing different types of  Hall effects.
For this we use the quasiclassical approach generalized in Refs.~\cite{Virtanen:2021,Virtanen:2022} to include SOC.

To describe the transport properties of the system we use the gauge covariant quasiclassical Green's function (GF) formalism. The GF $\check{g}(t,t')$ is an $8 \times 8$ matrix in Keldysh-Nambu-spin space~\cite{Belzig:1999,Feigelman:2000,Kamenev:2009}, $\check{g}=\begin{pmatrix}\check{g}^R & \check{g}^K\\ 0 & \check{g}^A\end{pmatrix}$, where the Keldysh GF $\check{g}^K=\check{g}^R \cdot \check{h} - \check{h} \cdot \check{g}^A$ describes the non-equilibrium properties of the system. Here, $\check{h}$ is the distribution function and the symbol $\cdot$ is used to denote a convolution in time, \textit{i.e.}, integration in the intermediate time variable. The $\check{.}$ symbol denotes matrices in Keldysh$\otimes$Nambu$\otimes$spin or Nambu$\otimes$spin space. The quasiclassical GF satisfies the normalization condition $\check{g} \cdot \check{g}=\delta(t-t')$.

In systems with time translational symmetry, the Green's function can be Fourier transformed in the $\tau=t-t'$ variable into the energy domain as
\begin{equation}\label{Fourier transform}
    \check{g}(\epsilon)=\int d\tau e^{i\epsilon \tau/\hbar}\check{g}(\tau)\; .
\end{equation}
For a bulk superconductor the retarded and advanced GFs in the energy domain are given by 
\begin{equation}\label{bulk GF}
    \check{g}^{R/A}_0(\epsilon)=g_0(\epsilon)\tau_3+f_0(\epsilon)\tau_1\; ,
\end{equation}
where $g_0$ and $f_0$ are the normal and anomalous parts of the bulk GF
\begin{subequations}\label{bulk g and f}
\begin{align}
    \label{bulk g}
    g_0(\epsilon)&=\frac{-i(\epsilon\pm i\Gamma)}{\sqrt{\Delta^2-(\epsilon\pm i\Gamma)^2}}\\
    \label{bulk f}
    f_0(\epsilon)&=\frac{\Delta}{\sqrt{\Delta^2-(\epsilon\pm i\Gamma)^2}} \; ,
\end{align}
\end{subequations}
and the equilibrium distribution function is given by $\check{h}(\epsilon)=\tanh{\frac{\epsilon}{2k_B T}}$. The upper and lower signs correspond to the retarded and advanced GFs respectively. The convergence factor $\Gamma \rightarrow 0^+$ guarantees that the retarded (advanced) GF is 0 for negative (positive) time $t-t'$. Nonetheless, a finite $\Gamma$ may also describe inelastic scattering effects present in real materials~\cite{Dynes:1978}. Such inelastic processes are responsible for the smoothing of the density of states peaks at the superconducting gap. The order parameter $\Delta(T)$ needs to be computed self-consistently~\cite{Kopnin:2001} with the gap equation 
\begin{equation}
\label{self-consistency}
\Delta\ln\left(\frac{T}{T_{c0}}\right)=2\pi k_B T\sum_{n=0}\left(f_0(\omega_n)-\frac{\Delta}{\omega_n}\right),
\end{equation}
where $\omega_n=2\pi k_B T(n+1/2)$, $n \in \mathbb{Z}$, are the Matsubara frequencies, 
$T_{c0}$ is the zero-field critical temperature and $f_0(\omega_n)$ is the Matsubara anomalous GF~\eqref{bulk f}, obtained by analytic continuation of the GF to the complex plane $\epsilon+i\Gamma \rightarrow i\omega_n$.

In diffusive systems where the scattering rate $\tau^{-1}$ is much higher than the other energy scales in the system, excluding the Fermi energy, the GF is determined from the well-known Usadel equation~\cite{Usadel}. The covariant version of the Usadel equation~\cite{Virtanen:2021,Virtanen:2022} allows describing the Hall and intrinsic spin Hall effects. For intrinsic SOC the Usadel equation reads~\cite{Bergeret:2014-SU2,Bergeret:2015-SU2,Tokatly:2017}
\begin{equation}\label{Usadel intrinsic}
    \hbar D\tilde{\nabla}_i\check{J}^i-\{\tau_3\hbar\partial_t,\check{g}\}-\left[i\check{A}^0\tau_3+\check{\Delta},\check{g}\right]=0\; ,
\end{equation}
and for extrinsic SOC~\cite{Virtanen:2021}
\begin{equation}\label{Usadel extrinsic}
    \hbar D(\tilde{\nabla}_i\check{J}^i+\check{\mathcal{T}})-\{\tau_3\hbar\partial_t,\check{g}\}-\left[i\check{A}^0\tau_3+\check{\Delta}+\frac{\sigma_i\check{g}\sigma_i}{8\tau_{\mathrm{SO}}},\check{g}\right]=0\; ,
\end{equation}
where $D$ is the diffusion coefficient, $\tau_{\mathrm{SO}}=9\tau/(8\lambda^4 p_F^4)$ is the spin-orbit relaxation time, $\lambda$ describes the SOC strength, $p_F$ is the Fermi momentum, $\tilde{\nabla}_i\check{X}=\partial_i\check{X} -i/\hbar[\check{A}^i,\check{X}]$ is the covariant derivative, $\check{\mathcal{T}}$ is an extrinsic SOC correction due to an effective torque originating from the spin Hall and the spin swapping effects~\cite{Bergeret:2016,Huang:2018}
\begin{equation}
    \check{\mathcal{T}}=i\varepsilon_{ijk} \frac{\varkappa}{4} [\tilde{\nabla}_i\check{g}\cdot\tilde{\nabla}_j\check{g},\sigma_k]+\varepsilon_{ijk}\frac{\theta}{4} [\sigma_k,\check{g}\cdot\tilde{\nabla}_i\check{g}\cdot\tilde{\nabla}_j\check{g}],
\end{equation}
and $\check{J}^i$ is the matrix current given by
\begin{equation}\label{matrix current}
    \check{J}^i=\check{g}\cdot\tilde{\nabla}_i\check{g}+\frac{\tau}{4m}(\{\check{F}^{ij}+\check{g}\cdot\check{F}^{ij}\cdot\check{g},\tilde{\nabla}_j\check{g}\}-i\hbar\tilde{\nabla}_j(\check{g}\cdot[\tilde{\nabla}_i\check{g},\tilde{\nabla}_j\check{g}]))
\end{equation}
for intrinsic SOC~\cite{Virtanen:2022} and
\begin{equation}\label{Extrinsic matrix current}
    \check{J}^i=\check{g}\cdot\tilde{\nabla}_i\check{g}-i\varepsilon_{ijk} \frac{\varkappa}{4} [\check{g}\cdot\tilde{\nabla}_j\check{g},\sigma_k+\check{g} \cdot \sigma_k \check{g}]-\varepsilon_{ijk} \frac{\theta}{4} \{\tilde{\nabla}_j\check{g},\sigma_k+\check{g} \cdot \sigma_k \check{g}\},
\end{equation}
for extrinsic SOC~\cite{Virtanen:2021}. Here $\varkappa=2p_F^2\lambda^2/3$ and $\theta=2\hbar p_F \lambda^2/\ell$ are spin-swapping and spin Hall coefficients~\cite{Lifshits:2009}, respectively, with $\varepsilon_{ijk}$ the Levi-Civita symbol and $\ell$ the mean-free path. The first term in Eqs.~\eqref{matrix current} and~\eqref{Extrinsic matrix current} is the standard diffusive current, while the second term is the leading contribution from spin-charge coupling describing the Hall effect.

The Usadel equation together with the normalization condition specifies the value of the GF. The observable quantities of the system are given by the GF, for example, the charge and spin currents are given by
\begin{equation}\label{charge current}
    j^{i0}(\boldsymbol{r},t)=-\frac{\pi\hbar\sigma_D}{8e}\mathrm{Tr}\left\{\tau_3\check{J}^i(\boldsymbol{r},t,t)^K\right\}
\end{equation}
\begin{equation}\label{spin current}
    j^{ij}(\boldsymbol{r},t)=\frac{\hbar}{2}\frac{\pi\hbar\sigma_D}{8e^2}\mathrm{Tr}\left\{\sigma_j\check{J}^i(\boldsymbol{r},t,t)^K\right\},
\end{equation}
where the Drude conductivity is given by $\sigma_D=\nu_F e^2 D$, $\nu_F$ is the density of states at the Fermi energy and the $K$ superscript denotes the Keldysh block of the matrix current.

\section{Onsager symmetries}\label{sec:onsager}
The Onsager reciprocal relations relate the conductivities between different pairs of driving fields and their conjugate currents. They demonstrate the reciprocity between inverse effects, such as the Hall, the spin Hall or spin-galvanic effects and their corresponding inverse effects.

The conductivity tensor in Eq.~\eqref{Ohms law} can be decomposed into four blocks,
\begin{equation}\label{eq:sigma}
    \sigma^{i\mu,j\nu}=\begin{pmatrix}
    \sigma^{i0,j0} & \sigma^{i0,jl}\\
    \sigma^{ik,j0} & \sigma^{ik,jl}
    \end{pmatrix}\; .
\end{equation}
The elements of the conductivity tensor are related through the Onsager reciprocal relations $\sigma^{j0,i0}(\boldsymbol{B})=\sigma^{i0,j0}(-\boldsymbol{B})$, $\sigma^{jl,ik}(\boldsymbol{B})=\sigma^{ik,jl}(-\boldsymbol{B})$ and $\sigma^{i0,jl}(\boldsymbol{B})=-\sigma^{jl,i0}(-\boldsymbol{B})$, where $\boldsymbol{B}$ comprises all time-reversal
symmetry (TRS) breaking fields~\cite{Jacquod:2012}. The minus sign in the last relation appears due to the spin currents having opposite T-parity to charge currents~\cite{Shi:2006}. The charge block $\sigma^{ij} \equiv \sigma^{i0,j0}$ is the usual $3 \times 3$ conductivity tensor describing the electric effects. The diagonal elements are the longitudinal conductivities (Ohm's law), while the off-diagonal elements describe the Hall effect. For instance, the system considered in Sec.~\ref{sec:hall} consists of a superconductor pierced by a magnetic field in the $z$ direction. The magnetic field breaks the TRS, allowing for non-zero transverse conductivities in the $xy$ plane. Due to rotational invariance around the $z$-axis, the transverse (Hall) conductivities are related through the relation $\sigma^{xy}=-\sigma^{yx}$. 

The spin block $\sigma^{ik,jl}$ is a $9 \times 9$ matrix relating the spin currents to the spin SU(2) fields. For instance, some of the off-diagonal elements of the spin block describe the spin-swapping effect. The spin-charge blocks $\sigma^{ik,j0}$ and $\sigma^{i0,jl}$ describe the spin Hall and inverse spin Hall effects, respectively. In Sec.~\ref{sec:spin Hall} we study the spin Hall effect in a superconductor due to intrinsic and extrinsic SOC. In the intrinsic case the SOC is of the Rashba type, while in the extrinsic case the SOC is introduced by impurities. Since there are no TRS breaking fields, the spin Hall conductivities satisfy $\sigma^{i0,jl}=-\sigma^{jl,i0}$. In the intrinsic case, inversion symmetry in the $z$ direction is broken by the Rashba SOC, so the spin Hall effect is restricted to the $xy$ plane. In the extrinsic case, the isotropy of the impurities results in spin currents in the plane perpendicular to the charge current direction $i$, so that the spin Hall conductivities are related through $\sigma^{lj,i0}=-\sigma^{jl,i0}$.

The susceptibility $\chi=i\omega\sigma$ is the response function to the vector potentials. From the fluctuation-dissipation theorem for $\chi$ \cite{LandauLifshitzVol5:1970}, it follows that the Hermitian part of the generalized conductivity tensor $\sigma'=\frac{1}{2}(\sigma+\sigma^\dagger)$ is the dissipative contribution, while the anti-Hermitian part $\sigma''=\frac{1}{2i}(\sigma-\sigma^\dagger)$ is the reactive contribution.

\section{Mattis-Bardeen theory within Quasiclassics}\label{sec:Mattis-Bardeen}
%
%
\begin{figure}[!t]
  \includegraphics[width=0.99\columnwidth]{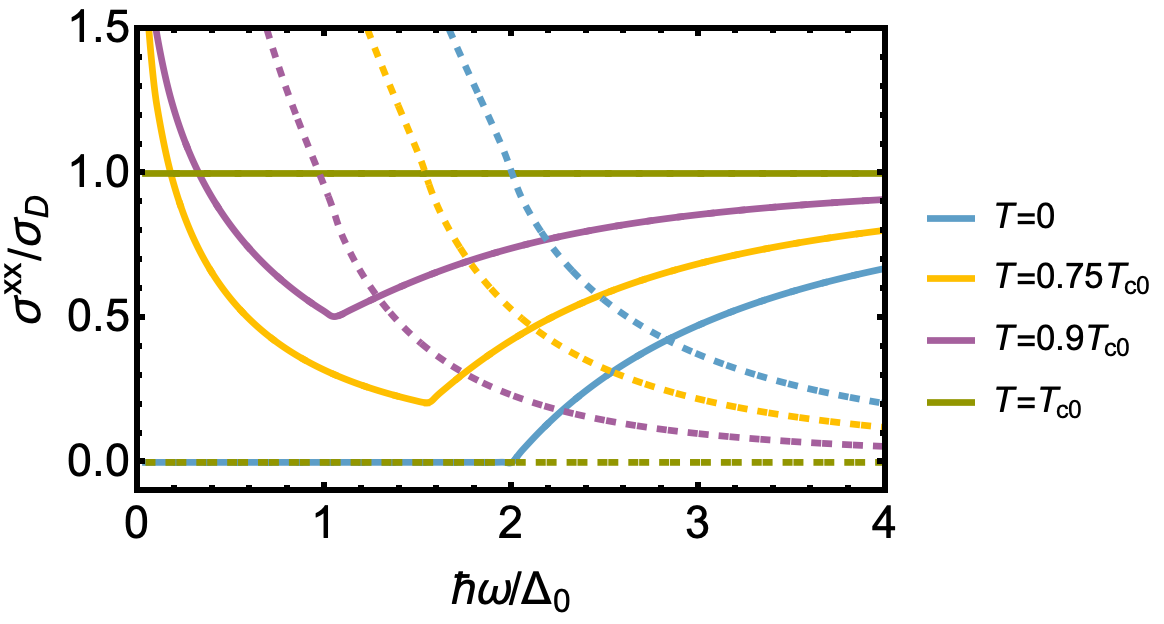}
\caption{Mattis-Bardeen response of a superconductor for different temperatures. The solid and dashed lines correspond to the real and imaginary parts of the conductivity, respectively.}\label{Fig:Mattis-Bardeen}
\end{figure}
%
%

In this section we apply the quasiclassical GF formalism introduced in Sec.~\ref{sec:theory} to the linear response theory to compute the longitudinal (charge) conductivity of a diffusive superconductor subjected to a time-dependent electric field $\boldsymbol{E}(t)$. In particular, we consider a superconductor in a microwave field which has extensively been experimentally realized~\cite{Glover:1956,Glover:1957,Sridhar:1988,Song:2009,Catto:2022}.

In diffusive normal metals the Drude model predicts a purely dissipative response~\cite{Ashcroft-Mermin}, that is to say, the current is in-phase with the electric field. Superconductors however show a reactive current which is most relevant at small frequencies compared to the superconducting gap. The frequency dependent microwave response of s-wave superconductors is described within the Mattis-Bardeen theory~\cite{mattis1958theory}. These results can be rederived in the present approach as follows.

We assume that the electric field is small enough so that it can be treated perturbatively. The bulk GF~\eqref{bulk GF} is corrected by the driving field, but the leading current contribution is given by the bulk GF. We start from a single-frequency electric field along the $x$-direction $\boldsymbol{E}(t)=E_0 e^{-i\omega t}\hat{\boldsymbol{u}}_x$ described via the vector potential $\boldsymbol{A}(t)=-iE_0/\omega e^{-i\omega t}\hat{\boldsymbol{u}}_x$. Therefore, the only non-zero term of the potential~\eqref{eq:fourpotential} is $\check{A}^x=-eA^x\tau_3$. Using Eq.~\eqref{charge current} we compute the longitudinal current $j^x(t)$, and dividing by the electric field $E_0 e^{-i\omega t}$ we obtain the longitudinal conductivity $\sigma^{xx}(\omega)$. The convolutions in Eq.~\eqref{matrix current} can conveniently be computed after a Fourier transformation into the energy domain. From~\eqref{charge current}, we get~\cite{supplemental}
\begin{widetext}
\begin{equation}\label{sigma_xx}
\begin{aligned}
    \sigma^{xx}(\omega)&=\frac{\sigma_D}{2\hbar\omega}\int d\epsilon\left[\Re{g_0(\epsilon_+)}\Re{g_0(\epsilon)}+\Im{f_0(\epsilon_+)}\Im{f_0(\epsilon)}\right][h(\epsilon_+)-h(\epsilon)]+i\{2[\Re{g_0(\epsilon)}\Im{g_0(\epsilon)}+\Re{f_0(\epsilon)}\Im{f_0(\epsilon)}]\\
    &h(\epsilon)+\left[-\Re{g_0(\epsilon_+)}\Im{g_0(\epsilon)}+\Im{f_0(\epsilon_+)}\Re{f_0(\epsilon)}\right]h(\epsilon_+)+\left[-\Im{g_0(\epsilon_+)}\Re{g_0(\epsilon)}+\Re{f_0(\epsilon_+)}\Im{f_0(\epsilon)}\right]h(\epsilon)\},\\
\end{aligned}
\end{equation}
\end{widetext}
where $\epsilon_+=\epsilon+\hbar\omega$. The expression is lengthy, but we can identify the different parts. The first term on the first line is the dissipative contribution. It yields the normal-state result $\sigma^{xx}=\sigma_D$ when $f_0(\epsilon)=0$ and $g_0(\epsilon)=1$. The rest of the terms are out of phase and describe the supercurrent effects. The prefactors of the distribution function terms are non-zero only when $\epsilon$ is of the order of $\Delta$ and decay at large energies, ensuring the convergence of the integrals. In general, the integral needs to be evaluated numerically.

%
%
\begin{figure*}[!t]
  \includegraphics[width=0.99\textwidth]{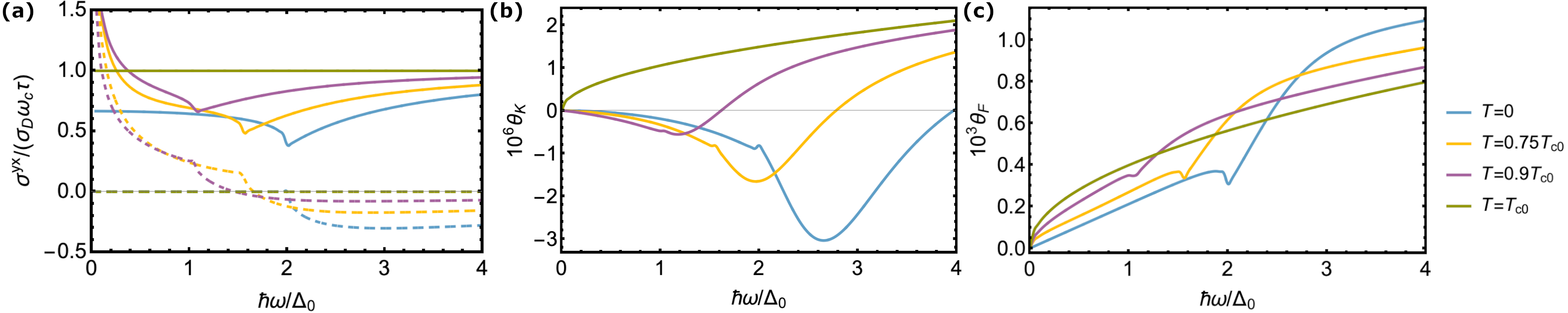}
\caption{(a) Hall response of a superconductor for different temperatures. The solid and dashed lines correspond to the real and imaginary parts of the conductivity, respectively. (b) Kerr and (c) Faraday rotation angles of linearly polarized incident light for different temperatures. The parameters used in panels (b) and (c) are $\tau=5 \times 10^{-2}\hbar/\Delta_0$, $\omega_c=0.2\Delta_0/\hbar$, $\omega_p=3 \times 10^4\Delta_0/\hbar$ and $d=0.2\lambda$, where $\lambda$ is the London penetration length.}\label{Fig:Hall}
\end{figure*}
%
%

In Fig.~\ref{Fig:Mattis-Bardeen} we show the frequency dependence of the longitudinal conductivity $\sigma^{xx}(\omega)$ for different temperatures~\cite{supplemental}. The conductivity is in agreement with the BCS theory and the experimental measurements~\cite{Glover:1956,Glover:1957}. At $T=0$ the real part of the conductivity is zero for $\hbar\omega<2\Delta_0$, where $\Delta_0=1.76 k_B T_{c0}$ is the zero-temperature gap, and it increases monotonously with a finite slope for $\hbar\omega>2\Delta_0$ so that it approaches the Drude conductivity in the $\omega\rightarrow\infty$ limit. At $T=0$ there are no thermally excited quasiparticles, so the processes that allow energy absorption are limited to the creation of electron-hole pairs, which require frequencies greater than $2\Delta_0$~\cite{Tinkham}. The effect of temperature in the conductivity is twofold; on the one hand, the superconducting gap $\Delta(T)$ is reduced with increasing temperature, so the the absorption edge is reduced to lower frequencies. On the other hand, at finite temperatures quasiparticles are thermally excited allowing energy absorption processes at lower frequencies.

Regarding the imaginary part of the conductivity, at $T=0$ it diverges as $1/\omega$ for $\hbar\omega \ll 2\Delta_0$. In superconductors where the electromagnetic field varies slowly in space on the scale of the coherence length the charge current is determined by the London equation~\cite{de_Gennes}. For diffusive superconductors at $T=0$, the London equation is given by $\boldsymbol{j}=-(\pi\Delta_0\sigma_D/\hbar)\boldsymbol{A}$, where the vector potential is given in the London gauge~\cite{Tinkham}. The London equation describes the free-acceleration aspect of the supercurrent response. For the plane wave vector potential considered here [see above Eq.~\eqref{sigma_xx}] the electric current is given by $\boldsymbol{j}=i\pi\Delta_0\sigma_D/(\hbar\omega)\boldsymbol{E}$, so at low frequencies the conductivity is purely imaginary and proportional to $1/\omega$.

At the critical temperature the gap is completely suppressed $\Delta(T_{c0})=0$ and the metal transitions into the normal state, so the conductivity is given by the AC Drude model~\cite{Ashcroft-Mermin}. In diffusive normal metals the Drude conductivity is frequency independent at frequencies much smaller than the elastic scattering rate and equal to the DC Drude conductivity $\sigma_D$.

\section{Ordinary Hall effect and Kerr rotation}
\label{sec:hall} 

%
%
\begin{figure*}[!t]
  \includegraphics[width=0.6\textwidth]{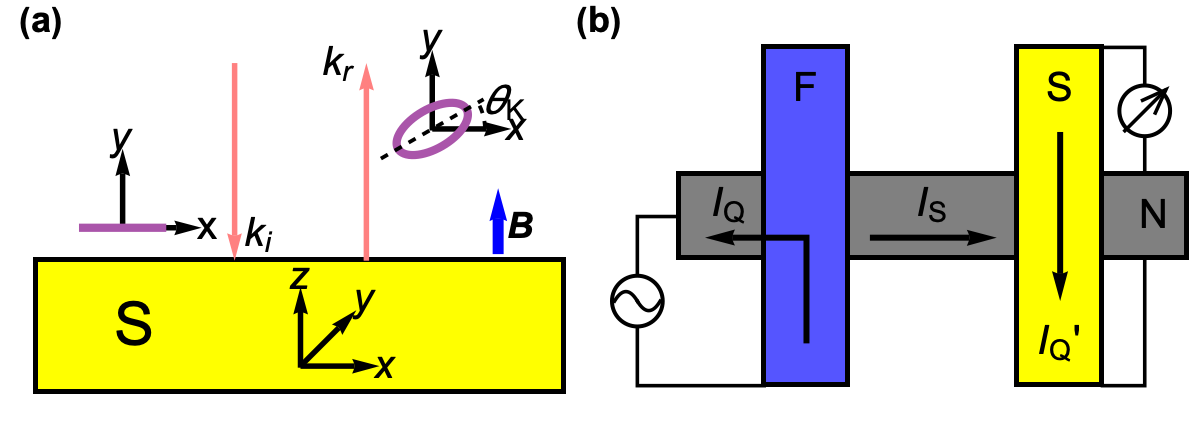}
\caption{(a) Proposed setup for the measurement of the Hall effect. Materials subjected to a magnetic field show circular birefringence, \emph{i.e.} left and right polarized light waves propagate with different velocities. The Kerr rotation angle is related to the longitudinal and transversal conductivities of the material. (b) Proposed setup for the detection of the inverse spin Hall effect in a superconductor using a lateral spin valve. If a charge current $I_Q$ is injected from the ferromagnet (F) to the normal metal (N), the nonequilibrium spin accumulation generated at the interface generates a pure spin current $I_\mathrm{S}$ to the right of F. The superconductor absorbs the spin current owing to its strong SOC, generating a charge current $I'_Q$ due to the inverse spin Hall effect.}\label{Fig:Hall_measurement}
\end{figure*}
%
%

In this section we study the Hall response of a superconductor in a microwave field subjected to a constant magnetic field $\boldsymbol{B}=B_0\hat{\boldsymbol{u}}_z$. In the Landau gauge, the vector potential is given by $\check{A}^x=-e(-iE_0/\omega e^{-i\omega t}-B_0 y)\tau_3$. The only non-zero elements of the field strength tensor~\eqref{field strength} are $\check{F}^{xy}=-\check{F}^{yx}=-eB_0\tau_3$. We assume that both the electric and magnetic fields are small and compute the electric currents to leading order in both fields. In addition to the longitudinal conductivity given by the Mattis-Bardeen response~\eqref{sigma_xx}, the system shows a transverse conductivity $\sigma^{yx}$ due to the interaction between the magnetic field and the electric current. The leading term of the Hall current is proportional to $E_0 B_0$ and it is given by the second current term in Eq.~\eqref{matrix current}. The Hall conductivity is given by the $y$ component of Eq.~\eqref{charge current}~\cite{supplemental}:
\begin{widetext}
\begin{equation}\label{sigma_yx}
\begin{aligned}
    \sigma^{yx}(\omega)=\frac{3\sigma_D \omega_c\tau}{4}+\frac{\sigma_D \omega_c\tau}{16\hbar\omega}\int & d\epsilon\; [g_0(\epsilon_+)K(\epsilon)^* +g_0(\epsilon)^*K(\epsilon_+)+2f_0(\epsilon)^* f_0(\epsilon_+)(g_0(\epsilon_+)+g_0(\epsilon)^*)](h(\epsilon_+)-h(\epsilon))\\
    &+[g_0(\epsilon_+)K(\epsilon)-g_0(\epsilon)K(\epsilon_+)+2f_0(\epsilon)f_0(\epsilon_+)(g_0(\epsilon_+)-g_0(\epsilon))]h(\epsilon)\\
    &+[g_0(\epsilon_+)^* K(\epsilon)^* -g_0(\epsilon)^*K(\epsilon_+)^*+2f_0(\epsilon)^*f_0(\epsilon_+)^*(g_0(\epsilon_+)^*-g_0(\epsilon)^*)]h(\epsilon_+),
\end{aligned}
\end{equation}
\end{widetext}
where $\omega_c=eB_0/m$ is the cyclotron frequency and $K(\epsilon)=g_0(\epsilon)^2-f_0(\epsilon)^2$. In the normal state $g_0(\epsilon)=1$ and $f_0(\epsilon)=0$, the first line in the integrand tends to $2[h(\epsilon_+)-h(\epsilon)]$ and integrates to $4\omega$, whereas the two other lines tend to zero. Together, these terms provide the normal-state Hall conductivity $\sigma^{yx} = \sigma_D \omega_c \tau$, valid to the first order in $\omega_c$.

In Fig.~\ref{Fig:Hall}(a) we show the Hall conductivity as a function of frequency at different temperatures. At $T=0$ the real part of the conductivity shows a sharp minimum at $\hbar\omega=2\Delta_0$ and asymptotically approaches the normal state Hall conductivity $\sigma^{yx}=\sigma_D\omega_c\tau$ at high frequencies. The real part of the conductivity is finite at $T=0$, while the imaginary part remains equal to zero until a threshold frequency is achieved. This is consistent with the experimental measurements of the Hall conductivity in superconducting YBa$_2$Cu$_3$O$_{7-\delta}$ samples by Spielman \textit{et al.}~\cite{Spielman:1994}.
As argued in Sec.~\ref{sec:onsager}, the dissipative contribution to the current is described by the Hermitian part of the conductivity tensor Eq.~\eqref{eq:sigma}. Due to the $\sigma^{xy}=-\sigma^{yx}$ symmetry relation, the dissipative part of the Hall current is given by the imaginary part of $\sigma^{yx}$, \textit{i.e.}, the out-of-phase component, while the reactive part is given by $\Re{\sigma^{yx}}$. This is the reason why at $T=0$ the imaginary part of the conductivity is non-zero only for $\hbar\omega>2\Delta_0$, so that the signal may be absorbed to create the electron-hole pairs. The temperature dependence of the Hall conductivity is very similar to that of the longitudinal conductivity discussed in Sec.~\ref{sec:Mattis-Bardeen}. The superconducting gap decreases with increasing temperature, so the absorption edge and the minimum of $\Re{\sigma^{yx}}$ are shifted to lower frequencies. For $T=T_{c0}$, we recover the normal state Hall conductivity. The normal-state Hall response is non-dissipative, in the leading order in $\omega_c\tau$, whereas in the superconducting state the response has a dissipative quasiparticle component.

Note that although the off-diagonal elements of $\sigma$ remain nonzero in the static limit $\omega \to 0$ in the superconducting state, all elements of the resistivity tensor $\rho=\sigma^{-1}$ vanish for $\omega \to 0$.
Hence, the relation $\boldsymbol{j}=\sigma \boldsymbol{E}$, or $\boldsymbol{E}=\rho \boldsymbol{j}$, does not imply here that uniform static supercurrent generates electric fields or a Hall effect.
Such equilibrium electric fields are expected to exist when the superflow is non-uniform, \cite{Bok:1968} but such configuration is not considered in the model here.

\subsection{Kerr and Faraday rotations}
The Hall conductivity can be probed optically through the Faraday or Kerr effect measurement as shown in Fig.~\ref{Fig:Hall_measurement}(a), where linearly polarized light transmitting or reflecting from the sample becomes elliptically polarized~\cite{Magneto-Optics,Ebert:1996}. The polarization rotation is described by the Faraday and Kerr angles which are in general complex quantities: their real part describes the amplitude of polarization rotation, and their imaginary part describes the ellipticity of the reflected polarization. These angles can be straightforwardly calculated from our theory. However, for simplicity the theory assumes a constant magnetic field, a situation that cannot be realized in the case of thick superconductors because of the Meissner effect expelling the field from inside the superconductor. However, the (Kerr) reflection is also a surface effect because of the finite skin depth of the electromagnetic field, and the transmission takes place only if the material is thinner than the corresponding skin depth. In other words, our estimates are accurate in the case where the London penetration depth is larger than either the skin depth (for Kerr reflection) or the sample thickness (for Faraday transmission). 

The skin depth can be obtained by solving the Maxwell equations inside the material. Disregarding the small correction from the Hall effect, it is obtained from
\begin{equation}\label{eq:skin_depth}
    \ell_{\rm skin}=\frac{c}{\omega {\rm Im} \sqrt{1+i\sigma^{xx}/{(\omega\varepsilon_0)}}}.
\end{equation}
On the other hand, the London penetration depth $\lambda$ is the zero-frequency limit of this skin depth. At $T=0$, $\sigma^{xx} \approx i\chi_0/\omega$, where $\chi_0=\pi\Delta_0\sigma_D/\hbar$, and hence $\lambda \approx c/\sqrt{\chi_0/\varepsilon_0}$.

For the simplest geometry, i.e., normal incidence of a linearly polarized electromagnetic wave onto the sample, using Maxwell equations and the boundary conditions, one can obtain the Kerr $\phi_K$ and Faraday $\phi_F$ angles as~\cite{Magneto-Optics,Ebert:1996}
\begin{equation}
\label{eq:kerrangle}
\phi_K=i\frac{r_+ - r_-}{r_+ + r_-} \simeq \frac{\sigma^{yx}}{\sigma^{xx} \sqrt{1+i\frac{\sigma^{xx}}{\omega\varepsilon_0}}},
\end{equation}
\begin{equation}
\label{eq:faradayangle}
\phi_F=\frac{\omega d}{c}\frac{r_+ - r_-}{(1-r_+)(1-r_-)} \simeq i\frac{d}{2c\varepsilon_0}\frac{\sigma^{yx}}{\sqrt{1+i\frac{\sigma^{xx}}{\omega\varepsilon_0}}},
\end{equation}
where $r_{\pm}$ are the reflection coefficients for left- and right-handed (with respect to the applied field) circularly polarized light, $\varepsilon_0$ is the vacuum permittivity and $d$ is the sample thickness. Note that the expression is obtained by assuming a small perturbation from the external magnetic field and only considering linear terms in the Hall conductivity ($\sigma^{yx}$). This approximation is valid when the Hall conductivity is much smaller than the longitudinal conductivity, a condition often met in many materials. Noting $\sigma_D$ as a natural scale of conductivity and defining the plasma frequency $\omega_p=\sqrt{\sigma_D/(\varepsilon_0 \tau)}$, we notice that the latter term inside the square root can also be written as $\sigma^{xx}/(\omega \varepsilon_0)=(\sigma^{xx}/\sigma_D) \omega_p^2 \tau/\omega$, providing a direct way to compare dimensionful quantities. For frequencies of interest here, $\omega \lesssim \Delta/\hbar$, the typical range is $\omega \ll \omega_p, 1/\tau$, and therefore the first term inside the square root in Eq.~\eqref{eq:kerrangle} can typically be disregarded. The order of magnitude of the polarization rotation is hence proportional to the small factor $\omega_c \tau \sqrt{\omega/(\omega_p^2 \tau)}$. 

The Faraday-Kerr rotation of the polarization state of light can experimentally be measured by passing the reflected light through a polarizer. The polarization direction is obtained by measuring the intensity of the reflected light with the polarizer oriented in parallel and perpendicular to the incident light. The Kerr (Faraday) rotation angle $\theta_{K(F)}$ specifies the rotation of the major axis of the elliptically polarized reflected light. It is given by the real part of $\phi_{K(F)}=\theta_{K(F)} + i \epsilon_{K(F)}$, plotted in Figs.~\ref{Fig:Hall}(b-c). The imaginary part $\epsilon_{K(F)}$ specifies the ratio of the minor to the major axes of the ellipsoid.

In Fig.~\ref{Fig:Hall}(b) we show the Kerr rotation angle~\eqref{eq:kerrangle} for normal incident light. The parameters used are $\tau=5 \times 10^{-2}\hbar/\Delta_0$, $\omega_c=0.2\Delta_0/\hbar$, and $\omega_p=3 \times 10^4\Delta_0/\hbar$; these  values are accessible in experiments. $\sigma^{xx}$ and $\sigma^{yx}$ are computed evaluating Eqs.~\eqref{sigma_xx} and \eqref{sigma_yx}. The Kerr rotation angle is of the order of $\mu$rad, which is an experimentally measurable rotation~\cite{Kato:2004}. In the normal state $\sigma^{xx}$ and $\sigma^{yx}$ are positive numbers, so $\theta_K$ is always positive. In the superconducting state both conductivities acquire an imaginary part, allowing for positive and negative values of $\theta_K$. In Fig.~\ref{Fig:Hall}(c) we show the Faraday rotation angle~\eqref{eq:faradayangle}. $\theta_F$ has a weaker dependence on temperature, but it is three orders of magnitude grater than $\theta_K$, so it is easier to measure than $\theta_K$.

Besides coupling to free-space light, the dynamical Hall effect can be accessed by studying the scattering parameters of microwaves in a multiterminal geometry. In particular, the matrix ${\mathcal S}$ of scattering parameters depends on the admittance matrix $Y(\omega)$ of the studied sample \cite{collin2001},
\begin{align}
   \mathcal{S}(\omega)
   =
   \frac{
      1 - \mathcal{Z}^{1/2} Y(\omega) \mathcal{Z}^{1/2}
   }{
     1 + \mathcal{Z}^{1/2} Y(\omega) \mathcal{Z}^{1/2}
   }
   \,,
\end{align}
where $\mathcal{Z}=\mathrm{diag}(Z_1,\ldots,Z_N)$ is a diagonal matrix containing the
characteristic impedances of transmission lines connected to each terminal $i$. This way, in case the bulk superconductor response gives the dominating contribution to the admittance matrix --- in other words, interface effects can be disregarded --- the Hall response can be related with the off-diagonal components of ${\mathcal S}$.

\section{Dynamical spin Hall response and its detection with magnetic resonance}
\label{sec:spin Hall}

In this section we study the spin Hall effect in a superconductor with SOC subjected to a microwave field. Several methods have been proposed and realized to measure the spin Hall and its inverse effects including electrical measurements~\cite{Valenzuela:2006,Takahashi:2012,Niimi:2014,Wakamura:2015} and Kerr rotation microscopy~\cite{Kato:2004}. In Fig.~\ref{Fig:Hall_measurement}(b) we propose a measurement setup based on a lateral spin valve. A lateral spin valve consists of a normal metal (N) bridging a ferromagnetic injector (F) and a detector, which in our case is a superconductor (S) with SOC. A charge current $I_Q$ is injected from F into the left side of $N$. The nonequilibrium spin accumulation generated at the interface is relaxed within the spin diffusion length, generating a pure spin current $I_S$ to the right of F. If the distance between the F and the S is shorter than the spin diffusion length, a nonequilibrium spin accumulation is generated at S~\cite{Niimi:2014}. The spin current is absorbed by the superconductor owing to its strong SOC. The polarization of the spin current is tuned to lie out-of-plane by applying a normal magnetic field. A perpendicular charge current $I'_Q$ is generated at the S due to the inverse spin Hall effect. This AC current can experimentally be measured by closing the S wire with a superconducting loop coupled to a rf-SQUID.

Alternatively, the measurement can be realized with a dynamic version of the setup used in Ref.~\cite{Jeon:2020}. There, two heavy-metal (Pt) injectors are used to generate and detect a magnon current in a ferromagnetic insulator. A heavy-metal superconductor placed in the middle absorbs part of the magnon current and converts it into a charge current via the inverse spin Hall effect. Replacing the DC injection by a finite-frequency injection then allows studying the AC spin Hall response of the superconductor. It also becomes interesting to separate the in- and out-of-phase oscillating parts of the detected signal, in comparison with the injected current.

\subsection{Intrinsic spin Hall response}\label{sec:intrinsic spin hall}

First, we study the spin Hall effect in a superconductor with Rashba SOC subjected to a microwave field. The Rashba SOC interaction is linear in momentum
\begin{equation}\label{Hamiltonian Rashba}
    \mathcal{H}_R=\alpha(\boldsymbol{p}\times\hat{\boldsymbol{u}}_z)\cdot\boldsymbol{\sigma}
\end{equation}
so it can be described through the SU(2) four-potential $\check{A}^x=m\alpha\sigma_y$, $\check{A}^y=-m\alpha\sigma_x$. The term $-eA^x\tau_3$ should be added to $\check{A}^x$ to account for the time-dependent electric field. For Rashba SOC the non-zero elements of the field strength tensor are $\check{F}^{xy}=-\check{F}^{yx}=2m^2 \alpha^2\sigma_z/\hbar$.

The terms contributing to the spin Hall current in Eq.~\eqref{matrix current} depend on the first-order correction of the GF due to the time-dependent electric field. We expand the GF to the first order in the electric field $\check{g}=\check{g}_0+\delta\check{g}+O(E_0^2)$, where $\delta\check{g}$ is the correction to the bulk GF. The Rashba SOC does not modify the bulk GF of a superconductor $\check{g}_0$, so it is given by Eqs.~\eqref{bulk GF} and~\eqref{bulk g and f}. Unlike the bulk GF, $\check{g}_0(t,t')=\check{g}_0(t-t')$, the first-order correction is not time-translation invariant due to the time-dependent electric field. Based on the time dependence of the electric field, we use the ansatz $\delta\check{g}(t,t')=e^{-i\omega t}\delta\check{g}(t-t')$ for the correction of the GF. The normalization condition for $\delta\check{g}$ reads $\check{g}_0(\epsilon_+)\delta\check{g}(\epsilon)+\delta\check{g}(\epsilon)\check{g}_0(\epsilon)=0$, and the Usadel equation~\eqref{Usadel intrinsic} to first order in $E_0$ is given by
\begin{multline}\label{Usadel correction}
    -\epsilon_\alpha([\sigma_y,\check{g}_0(\epsilon_+)[\sigma_y,\delta\check{g}(\epsilon)]]+[\sigma_x,\check{g}_0(\epsilon_+)[\sigma_x,\delta\check{g}(\epsilon)]])\\
    +i\hbar\omega\tau_3\delta\check{g}(\epsilon)+[i\epsilon\tau_3-\check{\Delta},\delta\check{g}(\epsilon)]\\
    =i\epsilon_E[\sigma_x,\{\sigma_z,\tau_3\check{g}_0(\epsilon)-\check{g}_0(\epsilon_+)\tau_3\}]\; ,
\end{multline}
where $\delta\check{g}(\epsilon)$ is the Fourier transform~\eqref{Fourier transform} of $\delta\check{g}(t-t')$, $\epsilon_\alpha=D m^2\alpha^2/\hbar$ is the Dyakonov-Perel energy (scattering rate) and $\epsilon_E=D\tau m^2\alpha^3 eE_0/(\hbar^2\omega)$. The spin structure of $\delta\check{g}(\epsilon)$ can be inferred from the Rashba Hamiltonian~\eqref{Hamiltonian Rashba}. As shown in Sec.~\ref{sec:Mattis-Bardeen}, the microwave field generates a longitudinal current along the $x$-direction. The Rashba SOC gives rise to intrinsic zero-field spin splitting~\cite{Bihlmayer:2006}. The motion of an electron in a two-dimensional electron gas through a perpendicular electric field results in a magnetic field in the rest frame of the electron that couples to the spin as given by Eq.~\eqref{Hamiltonian Rashba}, where the momentum-dependent magnetic field is $\mu_B \boldsymbol{B}_{\mathrm{eff}}=\alpha(\boldsymbol{p}\times\hat{\boldsymbol{u}}_z)$. Therefore, a vector potential in the $x$ direction spin splits the GF in the spin-$y$ direction as
$\delta\check{g}=\delta g_y\sigma_y\tau_3+\delta f_y\sigma_y\tau_1$. In Appendix~\ref{sec:GF correction} we solve Eq.~\eqref{Usadel correction} analytically and obtain closed form solutions for the retarded/advanced GFs $\delta\check{g}^{R/A}$ and the distribution function $\delta\check{h}$, where $\delta\check{g}^{K}(\epsilon)=\delta\check{g}^R(\epsilon)h(\epsilon)-h(\epsilon_+)\delta\check{g}^A(\epsilon)+\check{g}^R_0(\epsilon_+)\delta\check{h}(\epsilon)-\delta\check{h}(\epsilon)\check{g}^A_0(\epsilon)$.

%
%
\begin{figure}[!t]
  \includegraphics[width=0.99\columnwidth]{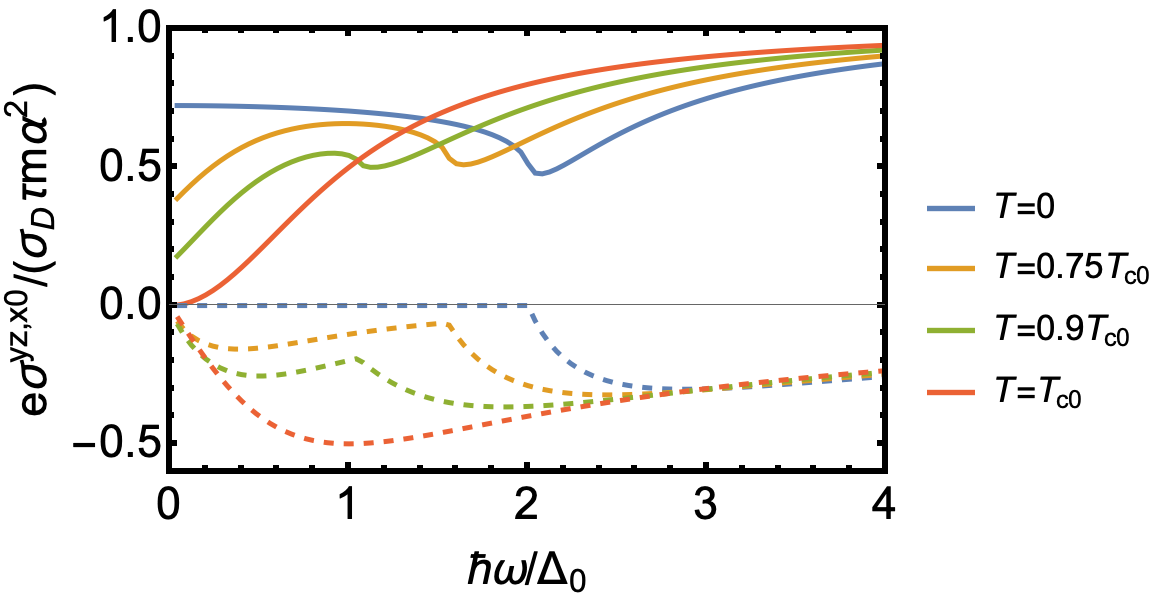}
\caption{Spin Hall response of a superconductor for different temperatures. The solid and dashed lines correspond to the real and imaginary parts of the conductivity, respectively. The value of the Dyakonov-Perel energy used is $\epsilon_\alpha=0.25\Delta_0$.}\label{Fig:Spin Hall}
\end{figure}
%
%

Plugging the solution into Eq.~\eqref{spin current} we obtain the spin Hall conductivity~\cite{supplemental}
\begin{widetext}
\begin{equation}\label{sigma_yzx0}
    \sigma^{yz,x0}(\omega)=\frac{\sigma_D \tau m\alpha^2}{e}-\frac{\sigma_D m\alpha}{4e^2 E_0}\int d\epsilon g_0(\epsilon_+)\delta g^K_y(\epsilon)+f_0(\epsilon_+)\delta f^K_y(\epsilon)+2[\Re{g_0(\epsilon_+)}\delta g^A_y(\epsilon)+i\Im{f_0(\epsilon_+)}\delta f^A_y(\epsilon)]h(\epsilon_+)\; .
\end{equation}
\end{widetext}

In Fig.~\ref{Fig:Spin Hall} we show the spin Hall conductivity for a superconductor with intrinsic SOC. The value of the Dyakonov-Perel energy used is $\epsilon_\alpha=0.25\Delta_0$. In Appendix~\ref{sec:normal state} we obtain a closed form expression for the spin Hall conductivity in the normal state, which agrees with literature predictions~\cite{Mishchenko:2004,Gorini:2012}. The spin conductivity depends on two characteristic frequency scales related to $\epsilon_\alpha$ and $\Delta(T)$. In homogeneous metals, the DC spin current is covariantly conserved unless extrinsic sources of spin-relaxation such as magnetic impurities are included~\cite{Sanz-Fernandez:2019}; or non-linear in $\boldsymbol{p}$ SOC is considered, such as cubic Dresselhaus interaction~\cite{Malshukov:2005}. This shows up as a vanishing spin Hall conductivity at $\omega=0$. The real part of the conductivity increases monotonically with an increasing frequency, reaching the asymptotic value $\sigma^{yz,x0}=\sigma_D \tau m\alpha^2/e$ for $\hbar\omega\gg\epsilon_\alpha$, while the imaginary part reaches an extremum at $\hbar\omega=4\epsilon_\alpha$ and decays to $0$ at high frequencies.

In the superconducting state, the temperature dependence of the spin Hall conductivity is most relevant at frequencies lower than $2\Delta$. The absolute values of the real and imaginary parts of the spin Hall conductivity have a minimum at $\hbar\omega=2\Delta$. Due to the $\sigma^{i0,jl}=-\sigma^{jl,i0}$ Onsager relation, the dissipative component of the spin Hall conductivity is given by the imaginary part of $\sigma^{yz,x0}$. Similar to the ordinary Hall response, at $T=0$ the out-of-phase spin current vanishes below the absorption edge $2\Delta_0$.

\subsection{Extrinsic spin Hall response}\label{sec:extrinsic spin hall}

We consider the response of the system with SOC due to extrinsic impurity scattering. In this case, the matrix current takes the form~\eqref{Extrinsic matrix current}. Due to the isotropy of the impurity scattering, we have spin Hall currents in both $y$ and $z$ directions. In this case the spin currents~\eqref{spin current} can be computed analytically. The conductivities are given by
\begin{equation}\label{extrinsic-sigma_zyx0}
\sigma^{zy,x0}=-\sigma^{yz,x0}=\frac{\hbar \sigma_D \theta}{2e},
\end{equation}
which do not depend on the frequency or temperature. This is a consequence of the diffusive regime considered in this work. For extrinsic SOC, the spin Hall conductivity is real, \emph{i.e.}, it is non-dissipative. In the following section we study the inverse Hall effect in systems with intrinsic and extrinsic SOC and explicitly show that the Onsager relations are satisfied.

\section{Inverse spin Hall response}
\label{sec:Inverse spin Hall response}
In this section we compute the charge current generated in systems with intrinsic and extrinsic SOC due to the inverse Hall effect. The charge current and the U(1) electric field are conjugate variables, in the same way the spin current has a conjugate force field $E^{jl}$ which generates $l$-polarized spin currents along the $j$ direction. This force field can be generated by the gradient of a Zeeman field, a time-dependent SOC or a spin dependent chemical potential~\cite{Zutic:2002,Fabian:2002}.

\subsection{Intrinsic SOC}
\label{sec:Intrinsic inverse spin Hall response}
We consider an SU(2) driving field $E^{yz}(t)=\mathcal{E}_0 e^{-i\omega t}$ which generates a $z$-polarized spin current in the $y$-direction. This electric field is described via the SU(2) vector potential $\check{A}^{y}(t)=-i\hbar\mathcal{E}_0/(2\omega) e^{-i\omega t}\sigma_z$. Taking the Rashba SOC into account, the vector potentials are given by $\check{A}^x=m\alpha\sigma_y$, $\check{A}^y=-m\alpha\sigma_x-i\hbar\mathcal{E}_0/(2\omega) e^{-i\omega t}\sigma_z$, so that the non-zero elements of the field strength tensor are $\check{F}^{xy}=-\check{F}^{yx}=2m^2 \alpha^2\sigma_z/\hbar-i (m\alpha \mathcal{E}_0/\omega) e^{-i\omega t}\sigma_x$. Following a similar procedure to the one used in Sec.~\ref{sec:intrinsic spin hall}, we obtain the correction to the bulk GF due to the driving field [see Appendix~\ref{sec:GF correction}] and compute the charge current along the $x$-direction using Eq.~\eqref{charge current}. The inverse spin Hall conductivity is
\begin{equation}\label{sigma_x0yz}
    \sigma^{x0,yz}(\omega)=-\frac{\sigma_D \tau m\alpha^2}{e}+\frac{\sigma_D \tau m^2\alpha^3}{e \hbar^2\mathcal{E}_0}\int d\epsilon \delta g^K_y(\epsilon).
\end{equation}
Evaluating Eqs.~\eqref{sigma_yzx0} and~\eqref{sigma_x0yz} numerically we have checked that the spin Hall and inverse spin Hall conductivities satisfy the Onsager relation introduced in Sec.~\ref{sec:onsager} $\sigma^{x0,yz}=-\sigma^{yz,x0}$.

\subsection{Extrinsic SOC}
As it has been argued in Sec.~\ref{sec:extrinsic spin hall}, systems with extrinsic SOC subjected to an electric field generate spin currents in both directions perpendicular to the electric field. For this reason, we consider two driving fields $E^{yz}(t)=\mathcal{E}_0 e^{-i\omega t}$ and $E^{zy}(t)=\mathcal{E}_0 e^{-i\omega t}$ and compute the charge current generated in the $x$-direction in each case. Following an equivalent procedure to Sec.~\ref{sec:extrinsic spin hall}, it is possible to obtain the inverse spin Hall conductivities analytically. They are
\begin{equation}\label{extrinsic-sigma_x0zy}
\sigma^{x0,zy}=-\sigma^{x0,yz}=-\frac{\hbar\sigma_D \theta}{2e}.
\end{equation}
In this case we may evaluate the Onsager symmetry analytically by comparing Eqs.~\eqref{extrinsic-sigma_zyx0} and~\eqref{extrinsic-sigma_x0zy} to obtain  $\sigma^{i0,jl}=-\sigma^{jl,i0}$. The proportionality between the conductivities shows that the spin Hall and inverse spin Hall effects are reciprocal effects.

\section{Conclusions}
\label{sec:conclusions}

In this work, we have used a unified description of charge and spin transport to study the dynamical response of dissipative superconductors to U(1) and SU(2) electric fields. We have used the gauge covariant quasiclassical GF formalism to obtain the charge and spin conductivities of superconductors in the presence of magnetic fields and spin-orbit interaction. Our model recovers known results in the appropriate limits, such as the normal state Hall conductivity and the spin Hall conductivity for normal metals with Rashba spin-orbit coupling. While diffusive normal metals show a purely dissipative response, superconductors show a reactive current that decays in frequency as $\omega^{-1}$, as described by the Mattis-Bardeen theory. We have analyzed the Onsager reciprocal relations between the direct and inverse Hall effects and have explicitly shown that they are satisfied.

Our findings show that both the ordinary and spin Hall conductivities show a dissipative component related to the out-of-phase current. In the case of the ordinary  Hall effect, the dissipative current contribution only arises in the superconducting state. For intrinsic spin Hall effect, the imaginary (dissipative) part of the Hall conductivity is always weaker in the superconducting state than in the normal state. At low frequencies, the spin Hall conductivity of a superconductor with Rashba SOC is dominated by the in-phase component, while in the normal state it is of the same order as the out-of-phase component. For extrinsic SOC, the spin Hall response is frequency and temperature independent and proportional to the spin Hall angle. In other words, there is no correction from superconductivity on the extrinsic spin Hall conductivity.

The dynamical Hall effect can be observed in optical spectroscopy via the Faraday-Kerr rotation of the polarization state of light in conventional superconductors. Suitable materials for the measurement of the spin Hall effect due to intrinsic SOC are Bi$_2$Se$_3$/monolayer NbSe$_2$ heterostructures~\cite{Yi:2022}, or LaAlO$_3$/SrTiO$_3$ interfaces~\cite{Caviglia:2010,BenShalom:2010}, where the Rashba SOC can be tuned by applying a gate voltage. For the extrinsic spin Hall effect we propose Nb~\cite{Jeon:2018}, NbN~\cite{Wakamura:2015}, and V~\cite{Wang:2017}, as they are superconductors with sizable impurity-induced SOC. For the detection of the spin Hall effect we propose a lateral spin valve, where a superconductor with SOC is used as a detector [see Fig.~\ref{Fig:Hall_measurement}(b)].

Our study focuses on investigating the dynamic charge and spin responses in conventional singlet single-band superconductors. It is worth noting  that spin ordering,  as for example  on iron based superconductors~\cite{Kamihara:2006,Kamihara:2008}, would be interesting to study. Our results can be readily generalized to include the effects of such spin ordering for example via the presence of an exchange field. Moreover, orbital degrees of freedom in multiband superconductors may provide additional dynamical channels, which possibly also show up in dynamic Hall-like responses, such as the valley Hall effect. Such effects may become visible in the dynamic responses of superconducting twisted multilayer graphene or field-biased bilayer graphene. Describing such effects would require generalizing our quasiclassical approach to the multiband case.

\begin{acknowledgments}
We thank Y. Lu for useful discussion. A.H. acknowledges funding from the University of the Basque Country (Project PIF20/05). S.V. acknowledges JYU Visiting Fellow Programme Grant 2023 (Registry No. 1643/13.00.05.00/2022) from the University of Jyväskylä, and support from National Science Center (NCN) Agreement No. UMO-2020/38/E/ST3/00418 and Project No. 2018/29/B/ST3/00937. This work was supported by the Academy of Finland (Contract No.~317118 and 321982). 
F.S.B. and A.H. acknowledge financial support from Spanish MCIN/AEI/ 10.13039/501100011033 through projects PID2020-114252GB-I00 (SPIRIT) and  TED2021-130292B-C42,
and the Basque Government through grant IT-1591-22. 

\end{acknowledgments}

\appendix
\numberwithin{equation}{section}
\renewcommand{\thesubsection}{\arabic{subsection}}
\section{Closed form solution for the corrected GF}\label{sec:GF correction}
Unlike for extrinsic SOC, the spin conductivity for intrinsic SOC depends on the correction of the GF due to the electric field $\delta\check{g}$. In this appendix we solve Eq.~\eqref{Usadel correction} analytically and provide a closed form solution for $\delta\check{g}$. As argued in Sec.~\ref{sec:intrinsic spin hall}, $\delta\check{g}$ has the following spin structure: $\delta\check{g}=\delta \check{g}_y\sigma_y$. The retarded and advanced parts of the Usadel equation~\eqref{Usadel correction} are simplified to
\begin{multline}\label{Usadel correction y}
    -4\epsilon_\alpha\check{g}_0(\epsilon_+)\delta\check{g}_y(\epsilon)+i\hbar\omega\tau_3\delta\check{g}_y(\epsilon)+[i\epsilon\tau_3-\check{\Delta},\delta\check{g}_y(\epsilon)]\\
    =4\epsilon_E(\tau_3\check{g}_0(\epsilon)-\check{g}_0(\epsilon_+)\tau_3)\; .
\end{multline}
The l.h.s. of the commutator in Eq.~\eqref{Usadel correction y} is proportional to the bulk GF $\check{g}_0(\epsilon)$ [see Eqs.~\eqref{bulk GF} and~\eqref{bulk g and f}]. Using the normalization condition $\check{g}_0(\epsilon_+)\delta\check{g}_y(\epsilon)+\delta\check{g}_y(\epsilon)\check{g}_0(\epsilon)=0$, Eq.~\eqref{Usadel correction y} becomes a matrix equation of the form $\check{A}\delta\check{g}_y=\check{B}$, such that $\delta\check{g}_y$ is given by
\begin{multline}
    \delta\check{g}_y(\epsilon) = 4\epsilon_E\Bigl(-4\epsilon_\alpha\check{g}_0(\epsilon_+)+i\hbar\omega\tau_3+i\epsilon\tau_3-\check{\Delta}\Bigr.\\
    \Bigl.-\sqrt{\Delta^2-\epsilon^2}\check{g}_0(\epsilon_+)\Bigr)^{-1}(\tau_3\check{g}_0(\epsilon)-\check{g}_0(\epsilon_+)\tau_3)\; .
\end{multline}

The Keldysh equation for the distribution function $\delta\check{h}=\delta h_y\sigma_y$ reads
\begin{multline}\label{Usadel correction h_y}
    (-4\epsilon_\alpha\check{g}_0^R(\epsilon_+)+i\hbar\omega\tau_3)(\check{g}_0^R(\epsilon_+)-\check{g}_0^A(\epsilon))\delta h_y(\epsilon)\\
    +[i\epsilon\tau_3-\check{\Delta},\check{g}_0^R(\epsilon_+)-\check{g}_0^A(\epsilon)]\delta h_y(\epsilon)\\
    =4\epsilon_E(\check{g}^R_0(\epsilon_+)\tau_3-\tau_3\check{g}^A_0(\epsilon))(h(\epsilon)-h(\epsilon_+))\; ,
\end{multline}
so $\delta h_y$ is given by
\begin{multline}
    \delta h_y(\epsilon)=4\epsilon_E\bigl((-4\epsilon_\alpha\check{g}_0^R(\epsilon_+)+i\hbar\omega\tau_3)(\check{g}_0^R(\epsilon_+)-\check{g}_0^A(\epsilon))\bigr.\\
    \bigl.+[i\epsilon\tau_3-\check{\Delta},\check{g}_0^R(\epsilon_+)-\check{g}_0^A(\epsilon)]\bigr)^{-1}(\check{g}^R_0(\epsilon_+)\tau_3-\tau_3\check{g}^A_0(\epsilon))\\
    (h(\epsilon)-h(\epsilon_+))\; .
\end{multline}

Following a similar procedure for the inverse spin Hall effect, the correction of the retarded and advanced GFs due to the SU(2) electric field considered in Sec.~\ref{sec:Intrinsic inverse spin Hall response} is given by
\begin{multline}
    \delta\check{g}_y(\epsilon) = \epsilon_\mathcal{E}\Bigl(-4\epsilon_\alpha\check{g}_0(\epsilon_+)+i\hbar\omega\tau_3+i\epsilon\tau_3-\check{\Delta}\Bigr.\\
    \Bigl.-\sqrt{\Delta^2-\epsilon^2}\check{g}_0(\epsilon_+)\Bigr)^{-1}(\check{g}_0(\epsilon_+)\check{g}_0(\epsilon)-1)\; ,
\end{multline}
where $\epsilon_\mathcal{E}=D m\alpha \mathcal{E}_0/\omega$ and the correction to the distribution function is
\begin{multline}
    \delta h_y(\epsilon)=\epsilon_\mathcal{E}\bigl((-4\epsilon_\alpha\check{g}_0^R(\epsilon_+)+i\hbar\omega\tau_3)(\check{g}_0^R(\epsilon_+)-\check{g}_0^A(\epsilon))\bigr.\\
    \bigl.+[i\epsilon\tau_3-\check{\Delta},\check{g}_0^R(\epsilon_+)-\check{g}_0^A(\epsilon)]\bigr)^{-1}(\check{g}^R_0(\epsilon_+)\check{g}^A_0(\epsilon)-1)\\
    (h(\epsilon_+)-h(\epsilon))\; .
\end{multline}

\section{Spin Hall conductivity in the normal state}\label{sec:normal state}
In the normal state ($T \geq T_{c0}$) the bulk GF is given by
\begin{equation}
    \check{g}^R_0(\epsilon)=\tau_3\; , \quad \check{g}^A_0(\epsilon)=-\tau_3\; , \quad \check{g}^K_0(\epsilon)=2\tau_3 h(\epsilon)\; .
\end{equation}
The r.h.s. of the retarded and advanced parts of Eq.~\eqref{Usadel correction} vanish, so the solution to the homogeneous equations is $\delta\check{g}^R=\delta\check{g}^A=0$, \emph{i.e.}, $\check{g}^{R/A}$ are not corrected by the electric field. Solving the Keldysh part of Eq.~\eqref{Usadel correction}, we obtain the correction to the distribution function
\begin{equation}
    \delta\check{h}=\frac{4\epsilon_E}{i\hbar\omega-4\epsilon_\alpha}(h(\epsilon)-h(\epsilon_+))\sigma_y\; .
\end{equation}
Finally, we perform the integral in Eq.~\eqref{sigma_yzx0} analytically and obtain the spin Hall conductivity in the normal state:
\begin{equation}
    \sigma^{yz,x0}(\omega)=\frac{\sigma_D \tau m\alpha^2\hbar\omega}{e(\hbar\omega+4i\epsilon_\alpha)}\; .
\end{equation}

Following the same procedure for the inverse spin Hall effect, the correction to the GF is $\delta\check{g}^R=\delta\check{g}^A=0$ and
\begin{equation}
    \delta\check{h}=\frac{\epsilon_\mathcal{E}}{i\hbar\omega-4\epsilon_\alpha}(h(\epsilon)-h(\epsilon_+))\sigma_y\; .
\end{equation}
The inverse spin Hall conductance Eq.~\eqref{sigma_x0yz} in the normal state is simplified to
\begin{equation}
    \sigma^{x0,yz}(\omega)=-\frac{\sigma_D \tau m\alpha^2\hbar\omega}{e(\hbar\omega+4i\epsilon_\alpha)}\; ,
\end{equation}
satisfying the Onsager relation $\sigma^{x0,yz}=-\sigma^{yz,x0}$.

%

\end{document}